\documentclass[10pt,aps,prl,floatfix,twocolumn,nofootinbib]{revtex4-2}

\usepackage[labelfont={small},subrefformat=parens,caption=false]{subfig}
\usepackage[dvipsnames]{xcolor}
\usepackage{amsfonts,amsmath,amssymb,bm}
\usepackage{graphicx}
\usepackage[utf8]{inputenc}
\usepackage{xspace}
\usepackage{isotope}
\usepackage{xparse}

\usepackage[pdfpagelabels, pdfencoding=auto, psdextra]{hyperref}
\hypersetup{%
 pdfsubject=Paper,
 pdfkeywords={nuclear physics} {Bayesian} {chiral EFT} {nuclear matter},
 unicode = true,
 breaklinks = true,
 colorlinks = true,
 linkcolor = blue,
 citecolor = blue,
 menucolor = blue,
 citecolor = blue,
 urlcolor = blue
}

\usepackage{orcidlink}

\graphicspath{{./}}

\usepackage{silence}
\newcommand{\cor}[1]{\textcolor{black}{#1}}
\newcommand\schro{Schr\"{o}dinger\xspace}
\newcommand{\adj}[1]{\tilde{#1}}

\newcommand{\so}{{S}}
\newcommand{\soadj}{\tilde{{S}}}
\newcommand{\psiadj}[1]{\tilde{\psi}_{#1}}

\newcommand{\wf}{\Psi}
\newcommand{\wfadj}{\tilde{\Psi}}

\newcommand{\matrixform}[1]{\begin{bmatrix} #1   \end{bmatrix}}

\newcommand{\resolventamp}{\mathcal{A}}

\newcommand{\vectheta}{\bm{\theta}}

\newcommand{\lamtwo}{\lambda}
\newcommand{\lamfour}{\lambda_4}

\newcommand{\perm}{\mathbb{P}}
\renewcommand{\Re}{\mathrm{Re}\,}
\renewcommand{\Im}{\mathrm{Im}\,}
\newcommand{\Neff}{N_\mathrm{eff}}

\newcommand\BCT{BCT\xspace}
\newcommand{\dimpara}{N_{p}}

\usepackage{amsmath,environ}
\NewEnviron{EqS}{%
\begin{equation}\begin{split}
  \BODY
\end{split}\end{equation}
}
\NewEnviron{EqSn}{%
\begin{equation*}\begin{split}
  \BODY
\end{split}\end{equation*}
}

\begin{document}

\title{Non-Hermitian Quantum Mechanics Approach for Extracting and Emulating Continuum Physics Based on Bound-State-Like Calculations} 

\author{Xilin Zhang~\orcidlink{0000-0001-9278-5359}}
\email{zhangx@frib.msu.edu}
\affiliation{\href{https://ror.org/03r4g9w46}{Facility for Rare Isotope Beams}, \href{https://ror.org/05hs6h993}{Michigan State University}, East Lansing, MI~48824, USA}

\date{\today}

\begin{abstract}
This work introduces a unified emulation framework for studying continuum physics in finite quantum systems. Using a reduced basis method, we construct powerful emulators for the inhomogeneous Schr\"{o}dinger equation that operate in a combined parameter space of complex energy ($E$) and other inputs ($\bm{\theta}$). Within the space, the emulators simultaneously perform analytical continuation in $E$---extracting continuum physics from numerically simpler bound-state-like calculations---and interpolate this entire process across $\bm{\theta}$. This yields a small, non-Hermitian system whose properties (e.g., resonances and scattering observables) can be rapidly predicted for any $\bm{\theta}$. Crucially, the complex-$E$ emulation provides a pathway to compute continuum observables for complex systems where advanced bound-state methods exist but direct continuum calculations are yet to be developed, while the $\bm{\theta}$-emulation enables rapid parameter-space exploration and can be adapted to accelerate other existing continuum calculations. Demonstrations with two- and three-body systems highlight the method's effectiveness and suggest its connection to (near-)optimal rational approximation. This Letter presents the key results, with further details reserved for a companion paper.
\end{abstract}

\maketitle

\paragraph{Introduction.}
This study concerns a finite quantum system, such as a nucleus, atom, or molecule, governed by the underlying Hamiltonian operator $H$. Of central interest is the matrix element of the resolvent operator between two spatially \emph{localized} sources at total energy $E$: 

\begin{align}
   \resolventamp(E,\vectheta) \equiv   \left\langle \tilde{S}(\vectheta) \left\vert\left[  M(E,\vectheta) \right]^{-1} \right\vert  S(\vectheta) \right\rangle  \ , \label{eq:resolventdef}
\end{align}
with $M(E,\vectheta) \equiv E - H(\vectheta)$, and $H$, $S$ and $\adj{S}$ depending on input $\vectheta$ (e.g., interaction couplings and kinematic variables). The $\vectheta$ dependencies are kept implicit below to make notations concise. $\resolventamp$ could be response functions, scattering amplitudes~\cite{GoldbergerQM,efros1985computation,Efros:2007nq,Zhang:2024gac}
\footnote{For computing the response function of a ground state $|\wf_\text{gs} \rangle $ induced by a transition operator $O$ at real $E$, the compact sources are $O| \wf_\text{gs}\rangle$. For scatterings, the sources could be compact as well~\cite{GoldbergerQM,efros1985computation,Efros:2007nq,Zhang:2024gac}; also see later discussion of a two-body system.}, 
or an essential part of computing optical potentials~\cite{Rotureau:2016jpf}. 
They are basically $ \langle \soadj \vert \wf\rangle $ or $ \langle \wfadj\vert \so \rangle $ with $\vert \wf\rangle $ and $\langle \wfadj\vert$ satisfying the inhomogeneous  \schro equations: 
  
\begin{align}
    M | \wf \rangle & = | \so\rangle\,  \ \text{and} \   \  \langle \wfadj | M    = \langle \soadj|  \ .   \label{eq:inhome}
\end{align} 

In the complex $E$ plane, $\resolventamp$ with a fixed $\vectheta$ has poles at locations given by the bound-state eigenenergies. Branch points (i.e., thresholds) also exist at which kinetic phase spaces for the system's fragmentation into subsystems start opening up; the associated branch cuts ({\BCT}s) are typically along the real-$E$ axis due to $H$'s Hermiticity~\cite{Zhang:2024gac}. 

If $H$ is approximated by a finite Hermitian matrix, $\resolventamp$'s {\BCT}s are discretized into poles located at the eigenenergies of the $H$-matrix---which are \emph{real}. This gives a poor approximation of $\resolventamp$ because it contradicts the fact that $\resolventamp (E)$ must be continuous when varying $E$ along the real axis with $\Im E $ fixed to $0^+$ or $0^-$; $\resolventamp$ could change rapidly there, but that is generally due to the resonance poles on adjacent Riemann sheets, not the unphysical \BCT poles.  

The type-I solution to this issue is to compute without discretizing the spectrum. It requires the calculations~\cite{Glockle:1983,Gloeckle:1995jg,DeltuvaCoulombReview2008,Deltuva:2012kt, Lazauskas:2019hil,Marcucci:2019hml,Navratil:2016ycn,Navratil:2022lvq, Descouvemont:2010cx,Nielsen:2001hbm} to handle wave functions' complicated oscillating asymptotics in coordinate space or singular functions in momentum space, when $E$ is real.   

The type-II methods, or non-Hermitian quantum mechanics (NHQM) approaches~\cite{ReinhardtComlexScaling2007, reinhardt1982complex, Moiseyev_2011}, move the {\BCT}s  below the real axis and discretize them~\cite{ReinhardtComlexScaling2007, reinhardt1982complex}~
\footnote{The \BCT poles don't necessarily line up along smooth \BCT curves. Still, they must be far enough away from the real axis to separate them from the resonance poles.}. 
Unphysical poles are  absent on the real axis. The resonance poles above the new {\BCT}s are now on the same Riemann sheet as the bound-state's, meaning both can be seen in $\resolventamp$ simultaneously. Per Eq.~\eqref{eq:resolventdef}, the resonances become $H$'s eigenstates. Methods of this type~
\footnote{Some of these methods are in fact connected~\cite{afnan1991resonances}.} include integration contour deformation~\cite{Glockle:1983}, complex scaling~\cite{reinhardt1982complex,Moiseyev_2011,Myo:2014ypa, Lazauskas:2011uj, Lazauskas:2012jc, Papadimitriou:2015rca, Lazauskas:2015ula, Lazauskas:2019hil}, and Berggren basis methods~\cite{Michel:2021jkx, Li:2021cqh, Berggren:1968zz, Michel:2002tm, Michel:2003jm, IdBetan:2002tv, Hagen:2006in, Rotureau:2006rf, Fossez:2017wpa, Hu:2020pxl, Michel:2022axa}. 

Another challenge is exploring $\resolventamp$ in the space of $\vectheta$ and $E$. It is usually infeasible to repeat the calculations many times. In fact, it may be unnecessary. According to the reduced basis methods (RBM)~\cite{hesthaven2015certified,Quarteroni:218966,Duguet:2023wuh,Melendez:2022kid,Drischler:2022ipa} in model order reduction (MOR)~\cite{Benner_2017aa,Benner2017modelRedApprox,benner2015survey}, the solution vector of an equation system, when varying a number ($\dimpara$) of its inputs, typically moves in a subspace with a low dimensionality (scaling mildly with $\dimpara$). Projecting the equations onto that subspace creates a smaller equation system, known as reduced-order model---a particular type of emulator or surrogate model which accurately and rapidly explores  the parameter space~\cite{Melendez:2022kid,Drischler:2022ipa, Duguet:2023wuh}. 
Such emulators are being studied for bound and resonance states~\cite{Frame:2017fah,Sarkar:2020mad,Sarkar:2021fpz,Konig:2019adq,Demol:2019yjt,Ekstrom:2019lss,Demol:2020mzd,Yoshida:2021jbl,Anderson:2022jhq,Giuliani:2022yna,Yapa:2023xyf} and continuum states~\footnote{Also see Ref.~\cite{Witala:2021xqm} for perturbation-based scattering emulators.}\cite{Furnstahl:2020abp,Drischler:2021qoy, Melendez:2021lyq,Zhang:2021jmi,Bai:2021xok, Drischler:2022yfb, Melendez:2022kid, Drischler:2022ipa, Bai:2022hjg, Garcia:2023slj, Odell:2023cun,Liu:2024pqp, Maldonado:2025ftg}. 

This work introduces a new NHQM approach that, via RBM-based subspace projections~\cite{Melendez:2022kid,Drischler:2022ipa,Duguet:2023wuh}, simultaneously (1) moves $\resolventamp(E,\vectheta)$'s {\BCT}s below the real-$E$ axis in the complex-$E$ plane and analytically continues $\resolventamp$ from above real-$E$ axis downward to the real-$E$ axis and further into the resonance region  and (2) emulates the analytical continuation process in the space of real $\vectheta$.

Specifically, at the first offline training stage, Eq.~\eqref{eq:inhome} is solved at a sample of $(E_\alpha^\mathrm{tr}, \vectheta_\alpha^\mathrm{tr})$ ($\alpha = 1,,,N_b$) with $\Im E_\alpha^\mathrm{tr} > 0$. This step is the only computationally intensive component of the approach. 
The solutions $| \wf(E_\alpha^\mathrm{tr}, \vectheta_\alpha^\mathrm{tr}) \rangle$ and $\langle \wfadj(E_\alpha^\mathrm{tr}, \vectheta_\alpha^\mathrm{tr}) |$, or simply $ | \wf_\alpha^\mathrm{tr} \rangle$  and $ \langle \wfadj_\alpha^\mathrm{tr} |$, satisfy bound-state-like boundary conditions~\cite{efros1985computation,Efros:2007nq}---which are much simpler to deal with numerically than the oscillating asymptotics in the real-$E$ cases. At the next online emulation step, $ | \wf_\alpha^\mathrm{tr} \rangle$  and $ \langle \wfadj_\alpha^\mathrm{tr} |$ are used as a subspace basis to construct general solutions: 

\begin{align}
|\wf \rangle &  = c_\alpha   | \wf_\alpha^\mathrm{tr} \rangle  \,   \ \text{and}  \  \
\langle  \wfadj |  =  \adj{c}_\alpha  \langle \wfadj^\mathrm{tr}_\alpha | \ .   \label{eq:RBMansatz}
\end{align}
The convention of summing over repeated indices is used. In this work, for simplicity, we typically over-sample the training points, though a singular value decomposition (SVD) of the norm matrix [defined in Eq.~\eqref{eq:eigproblem}] reveals a much smaller effective rank ($\Neff \ll N_b$). This redundancy points to future optimizations with greedy algorithms that pick training points in a more  strategic and efficient way ~\cite{Sarkar:2021fpz,Maldonado:2025ftg,Quarteroni:218966}.
 
To get $c_\alpha$ and $\adj{c}_\alpha$ as functions of $E$ and $\vectheta$, we plug the ansatz in Eq.~\eqref{eq:RBMansatz} into a variational approach~\cite{Pomraning1965} for solving linear equations and get a small $N_b$-dimension linear system~\cite{Zhang:2024gac}: with $\matrixform{M}_{\alpha\beta}  \equiv \langle \wfadj^\mathrm{tr}_\alpha | M | \wf^\mathrm{tr}_\beta \rangle$,

\begin{align}
    \matrixform{M}_{\alpha\beta} c_\beta & = \langle \wfadj^\mathrm{tr}_\alpha | \so \rangle \,   \ \text{and} \  \
     \adj{c}_\beta \matrixform{M}_{\beta\alpha}  = \langle \soadj | \wf^\mathrm{tr}_\alpha  \rangle \ . \label{eq:lineareq}   
\end{align}
The coefficients in these equations (e.g.,~$\matrixform{M}$) can be rapidly emulated when the parameter dependence in $M$ and $\so$ and $\soadj$ are factorized from the operators exactly (e.g., $E$ in $M$) or approximately~\footnote{Detailed discussion on this point can be found in e.g., Refs.~\cite{Drischler:2022ipa,Odell:2023cun}, concerning affine vs non-affine parameters.}.

$\resolventamp$ can then be emulated for general $E$ and $\vectheta$ values via $\matrixform{\resolventamp} =  c_\alpha \langle \soadj | \wf^\mathrm{tr}_\alpha \rangle = \adj{c}_\alpha \langle \wfadj^\mathrm{tr}_\alpha | \so\rangle$. Note 
that the approach's reliance on bound-state-like calculations provides a key strategic advantage~\cite{Moiseyev_2011,Efros:2007nq, Carbonell:2013ywa, Orlandini:2013eya}: leveraging advanced bound-state methods---which are generally applicable to larger systems than their continuum counterparts---to expand the reach of continuum calculations. In short, \emph{we now have fast \& accurate access to continuum physics, e.g., $\resolventamp(E,\theta)$, at different real values of $E$ and $\theta$, based on bound-state-like calculations.} 

In the following sections, we demonstrate the method in both two- and three-body systems—--importantly, the approach applies to general systems. Multiple ways of varying $E$ and or $\vectheta$ will be used to  realize the method's different functionalities. Then, the discussion section offers a broad perspective about the method, including its connections with the existing complex energy (CE)~\cite{Schlessinger:1966zz, Schlessinger:1968vsk, Schlessinger:1968zz, McDonald:1969zza, Uzu:2003ms, Deltuva:2012fa, Deltuva:2013qf,Deltuva:2013mda,Deltuva:2014pda} and Lorentz integral transformation (LIT)~\cite{efros1985computation, Efros:1994iq, Efros:2007nq,Orlandini:2013eya,Sobczyk:2021dwm,Sobczyk:2023sxh,Bonaiti:2024fft} methods. Afterward, future studies are discussed, followed by a brief summary. A detailed description of the work is provided in Ref.~\cite{Zhang:2024gac}. Source codes for generating the results of this work can be accessed via the companion website~\cite{BUQEYEsoftware}.

\begin{figure}
    \centering
    \includegraphics{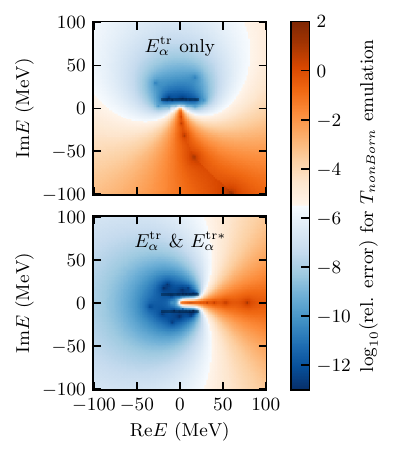}
    \caption{The relative error (in its logarithm) of an emulated resolvent matrix element for a two-body system in the complex-$E$ plane. The training energies are evenly separated on the black solid line(s) which are 10 MeV away from the real axis. }
    \label{fig:T-2dimerror-2b}
\end{figure}

\paragraph{Two-body demonstration.} We study a two-nucleon-like system in their $s$-wave channel, with a short-ranged interaction $V$ (see Ref.~\cite{Zhang:2024gac} for more details). Particle mass is $940$ MeV in natural units. Let $|p_{in}\rangle$ be a plane wave state, and $|\so\rangle$ and $|\soadj\rangle$ as $ V|p_{in}\rangle$. We fix $H$ and sources (including $p_{in}$) and vary $E$. With such sources, $\resolventamp$ becomes the non-Born term in the scattering $T$-matrix~\cite{efros1985computation,Zhang:2024gac}, but $-\Im\resolventamp$ also mimics response functions. Note that all the results of this work are based on training calculations with tiny errors ($\approx 10^{-12}$), allowing us to derive essential understandings of the approach. 

Figure~\ref{fig:T-2dimerror-2b} demonstrates the NHQM nature of the method, in particular, its dependence on $E_\alpha^\mathrm{tr}$'s locations in the complex-$E$ plane. 
It shows the emulation errors for $\matrixform{\resolventamp}$ in the complex plane. In the top panel, $N_b = 23$ training energies were evenly spaced along the solid black line with $\Im E^\mathrm{tr}_\alpha = 10$ MeV (note that the effective basis size $\Neff \approx 10$, see Ref.~\cite{Zhang:2024gac} for more details). The error is the smallest ($\approx$ the training calculation errors) in the blue region, close to the training energies.  When extrapolating, the error increases and diverges to infinity at the poles of $\matrixform{\resolventamp}$---see the dark orange dots. The curve formed by those pole dots can be considered as $\matrixform{\resolventamp}$'s discretized \BCT originating from the $E=0$ branch point.  Therefore,  the \BCT's location---in this case, below the real axis---is directly connected to the emulation error pattern; the latter is controlled by the locations of $E_\alpha^\mathrm{tr}$.

This connection is also seen in the bottom panel, where ${E^\mathrm{tr}_\alpha}^\ast$ are included in the training energies. There, the error pattern has a mirror symmetry with respect to the real axis, enforced by the symmetry of the training energy locations. As a result, the \BCT is back on the real axis. The contrast between the two panels suggests that breaking the mirror symmetry of training energy locations with respect to the real axis forces the \BCT away from the real axis, which, as discussed previously, is the basic feature of the NHQM methods.

To understand this point further, note that $\matrixform{\resolventamp}$'s poles are the poles of $c_\alpha$ and $\adj{c}_\alpha$. Their locations are given by the solutions of the generalized eigenvalue problem with  projected $H$ and norm matrices: 

\begin{align}
\matrixform{H}_{\alpha\beta} = \langle \tilde{\wf}_\alpha^\mathrm{tr} | H | \wf_\beta^\mathrm{tr} \rangle  \ \text{and} \  \matrixform{N}_{\alpha\beta} = \langle \tilde{\wf}_\alpha^\mathrm{tr} |  \wf_\beta^\mathrm{tr} \rangle    \ . \label{eq:eigproblem}
\end{align}
Both are non-Hermitian generally, even with a Hermitian $H$ in the training calculations; they are complex symmetric if $\so(\vectheta_\alpha)$ and $\soadj(\vectheta_\alpha)$ are equal and invariant under time-reversal transformation.
$\matrixform{H}$'s eigenenergies are thus complex, and  $\matrixform{\resolventamp}$'s {\BCT}s are off the real axis. But, as an exception, for the bottom panel of Fig.~\ref{fig:T-2dimerror-2b}, the eigenvalue problem turns Hermitian with only real eigenenergies. See Ref.~\cite{Zhang:2024gac} for  detailed discussions.

\begin{figure}
    \centering
    \includegraphics{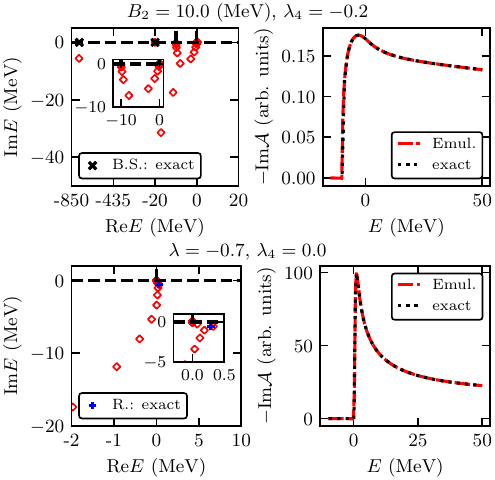}
    \caption{Emulation in $E$ at fixed $\vectheta$ for a three-body system with (top rows) and without (bottom rows) a bound dimer. The left panels show $\matrixform{\resolventamp}$'s eigenvalues ($\diamond$). Note the different scales on the two sides of the x-axis. The exact locations of bound or resonance states are also marked (see the legends). The insets zoom in on the regions around the branch points. The right panels compare the emulated and the exact calculations of $-\Im \resolventamp$ at real energies.}
    \label{fig:spec_3b_emul-E}
\end{figure}

\paragraph{Three-body demonstrations.}
We illustrate our method, including its $\vectheta$-emulation, for a $s$-wave interacting three-boson system studied previously in  Ref.~\cite{Zhang:2021jmi}, by emulating the solutions of Faddeev equations~\cite{Glockle:1983}. Emulation in terms of the \schro equation will be studied in the future. The equations can be cast into the form of Eq.~\eqref{eq:inhome}, with  

\begin{align}
  |\wf \rangle \equiv 
 \begin{pmatrix}
 | \psi_1 \rangle    \\
 | \psi_4 \rangle  
\end{pmatrix}      \   \mathrm{and} \ 
 |\wfadj \rangle \equiv 
\begin{pmatrix}
 | \psiadj{1} \rangle    \\
 | \psiadj{4} \rangle  
\end{pmatrix}     \ , 
\end{align}
and two-component  $|\so\rangle $ and $ |\soadj\rangle$. The sources are spatially localized and chosen so that $\langle \soadj | \wf \rangle$ and $\langle \wfadj |  \so \rangle$ are the non-Born term in the particle-dimer scattering $T$-matrix~\cite{Zhang:2024gac}. With $H_0$  the kinetic energy operator, $V_{1}$ the pair-wise interactions with a coupling strength $\lamtwo$, $V_4$  the  three-body interaction with a strength $\lamfour$, and $\perm$  the permutation operator~\cite{Glockle:1983},  
$M$ is 

\begin{align} 
\begin{pmatrix}
 E - H_0 - V_1 - V_1 \perm  & - V_1   \\
  -3 V_4 & E- H_0 - V_4   
\end{pmatrix}  \equiv  E - H  \ .
\end{align}

Below, we (1) vary $E$ alone to compute spectrum (i.e., eigenenergies) and $\resolventamp$ for a fixed $H$ and sources (as done in the two-body demonstration); (2) alter $E$ and the $\vectheta$ inside $H(\vectheta)$ to emulate the spectra and $\resolventamp$ for different Hamiltonians; and (3) vary $E$ and the $\vectheta$ in both $H$ and sources to gain the full set of functionalities, e.g., computing scattering amplitudes and emulating them in $\vectheta$.

We first extract the spectrum  of  $H(\vectheta)$ by emulating in $E$ while fixing $\vectheta$ in $H$ and the sources. Figure~\ref{fig:spec_3b_emul-E} shows the spectra ($\diamond$ in left panels) and the related $-\Im \resolventamp$ (right panels) in two cases: the $H$ in the top row has a bound dimer with binding energy $B_2 = 10$ MeV, and the bottom has no bound dimers but a three-body resonance. The associated $\lamfour$ and $\lamtwo$ (or $B_2$) are in the titles. Both emulators have the same sources and  $N_b = 48$ training energies with $\Im E_\alpha^\mathrm{tr} = 3$ MeV and $\Re E_\alpha^\mathrm{tr}\in [-20, 50]$ MeV ($\Neff \approx 15$ for both cases).  

The left panels in Fig.~\ref{fig:spec_3b_emul-E} again demonstrate the NHQM nature of the method: most eigenenergies form off-axis curves, representing $\matrixform{\resolventamp}$'s discretized {\BCT}s which start from the branch point(s) on the real axis  (exact values marked by black vertical lines).  In the top left panel, there are two branch points corresponding to particle-dimer and three-particle thresholds, while only a single branch point exists in the bottom left. 

The energies of the physical states (as the subsets of all $\diamond$s), including the three-body bound states in the top left panel and the resonance in the bottom left, agree well with the exact results, marked as ``$\times$'' and ``$+$'' respectively. Note that broad resonances are more difficult to extract than narrow ones, as they are located further away from the region of the training calculations, where larger extrapolation error is expected~\cite{Zhang:2024gac}. 

Importantly, the pattern in which the \BCT states and isolated bound or resonance states are distributed in Fig.~\ref{fig:spec_3b_emul-E} helps us identify the physical states~\cite{Zhang:2024gac}, considering that we can't order the complex eigenenergies to identify states as done in dealing with real spectra. 
Moreover, the \BCT poles are exponentially clustered towards the branch point(s) in Figs.~\ref{fig:spec_3b_emul-E} (and~\ref{fig:T-2dimerror-2b}), mirroring that found in recent studies of \mbox{(near-)optimal} rational approximation (ORA) for functions with branch points~\cite{Nakatsukasa_2018, Trefethen2021NodeClustering, Trefethen2023review, nakatsukasa2023years}. Such similarities~\cite{Zhang:2024gac} suggest a connection between the two methods. The robustness of ORA~\cite{Nakatsukasa_2018, nakatsukasa2023years} thus lends support to our complex-$E$ emulation as an advanced analytical continuation tool.

Finally, in the right panels of Fig.~\ref{fig:spec_3b_emul-E}, the emulated and exact calculations of $-\Im\resolventamp$  at real energies 
are very close, including near thresholds. Note that $-\Im\resolventamp$ mimics  typical response functions. 

\begin{figure}
    \centering
    \includegraphics{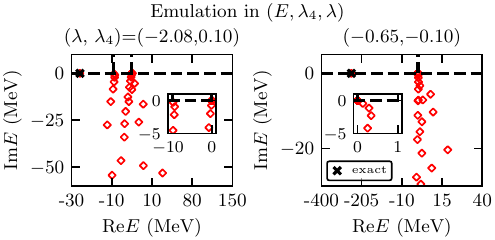}
    \caption{Emulated spectra (marked as $\diamond$s) by a single emulator. The insets zoom in on the regions around the branch points. Two testing results are plotted, with the parameter values shown in the panel titles. The exact locations of the physical states (no resonance here) are marked by ``$\times$'' in both panels. }
    \label{fig:spec_3b_emul-EAndlams}
\end{figure}
Now, can we emulate the spectrum calculations for different $H$? Yes, if we  vary $E$ and the $\vectheta$ inside $H(\vectheta)$. Here, a single emulator is trained to emulate in $E$,  $\lamfour$ and $\lamtwo$.  We fix the sources and $\Im E_\alpha^\mathrm{tr}$ to be the same as in Fig.~\ref{fig:spec_3b_emul-E}, and then  sample $N_b = 60$ training points ($\Neff \approx 43$) in the $(\Re E_\alpha^\mathrm{tr}, \lamfour, \lamtwo)$ space using Latin hypercube sampling (LHS)~\cite{doi:10.1080/01621459.1993.10476423} with $\Re E_\alpha^\mathrm{tr} \in [-20, 50]$ MeV, $\lamfour \in [-0.5, 0.5]$, and  $\lamtwo \in [-2.15, 1]$. The emulator is then tested at two instances, as shown in Fig.~\ref{fig:spec_3b_emul-EAndlams}, corresponding to two distinctive systems: one has a bound dimer in the left, while none in the right (see the panel titles for the parameter values).

These emulated spectra are qualitatively similar to those in Fig.~\ref{fig:spec_3b_emul-E}, including the agreement between the emulated physical states and the exact results. The results thus demonstrate the spectrum emulation capability of our method.  However, deep into the complex plane (10s MeV below the real axis), the supposedly \BCT eigenvalues are scattered somewhat. This is not problematic for emulating real-$E$ observables and near-axis resonances but can make it difficult to separate possible broad resonancs from the nearby  states that supposedly represent \BCT(s). This phenomenon needs further studies.

\begin{figure}
    \centering \includegraphics{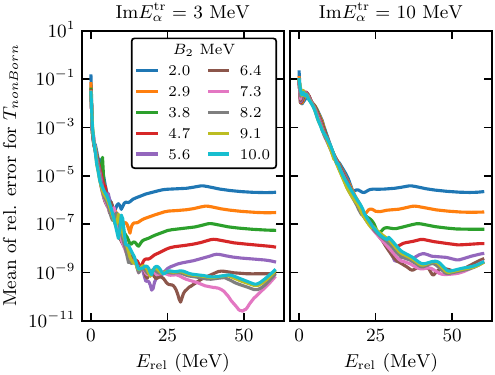}
    \caption{The non-Born term of the particle-dimer scattering $T$-matrix are emulated by two  emulators with different $\Im E^\mathrm{tr}_\alpha$. For each curve, $\lamtwo$ and thus $B_2$  are fixed  (see the legend), but a sample of $\lamfour$ values is chosen to test the emulators. The means of the sample of relative errors are plotted.}
    \label{fig:Temul1dimerror_3b}
\end{figure}

To compute and emulate particle-dimer scattering amplitudes, we must vary the source parameters, including the interaction couplings and the scattering energy $E_\mathrm{rel}$ between the particle and dimer, in addition to those in $H$. For on-shell scatterings, as computed in the emulation stage, the real-valued $E_\mathrm{rel}$ is related to $E$ via $E_\mathrm{rel} - B_2 = E$. However, the training calculations treat them as separate emulation variables. Thus, we sample $N_b =128$ training points using LHS in the $(\Re E, E_\mathrm{rel}, \lamtwo, \lamfour)$ space, with 
$E_\mathrm{rel} \in [0, 60]$ MeV, $\lamtwo$ in a range in which $B_2 \in [2, 10]$ MeV, and  $\Re E_\alpha^\mathrm{tr}$ and $\lamfour$ in the same ranges as for Fig.~\ref{fig:spec_3b_emul-EAndlams}. 

Two different emulators with $\Im E^\mathrm{tr}_\alpha = 3$ and $10$ are trained. They are checked against the exact on-shell \cor{$T$-matrix non-Born terms} at a sample of 1000 points in the $(E_\mathrm{rel}, \lamtwo, \lamfour)$ space. \cor{In Fig.~\ref{fig:Temul1dimerror_3b},} each curve represents the mean of the relative emulation errors from a sub-group of testing points with the same $B_2$ but different $\lamfour$.  \cor{The tiny errors at energies above thresholds suggest robust emulation performances there. Meanwhile, the (relative) } errors increase towards the particle-dimer threshold, similar to the behaviors of the rational approximations of functions near their branch points. \cor{However, reducing $\Im E^\mathrm{tr}_\alpha$ (from $10$ to $3$ MeV here) improves the emulations around thresholds systemically.}. 

\paragraph{Discussion.}
Our emulators are trained using the calculations performed with complex $E_\alpha^\mathrm{tr}$ and real $\vectheta_\alpha^\mathrm{tr}$. The cost of individual training  calculation increases with system size, but its feasibility for few- and many-body systems have been demonstrated in previous studies~\cite{Carbonell:2013ywa} employing  the CE~\cite{Schlessinger:1966zz, Schlessinger:1968vsk, Schlessinger:1968zz, McDonald:1969zza, Uzu:2003ms, Deltuva:2012fa, Deltuva:2013qf,Deltuva:2013mda,Deltuva:2014pda}~\footnote{\cor{Here, training points typically differ only by their $\Im E^\mathrm{tr}_\alpha$ values.}} and LIT methods~\cite{efros1985computation, Efros:1994iq, Efros:2007nq,Orlandini:2013eya,Sobczyk:2021dwm,Sobczyk:2023sxh,Bonaiti:2024fft}. These works underscore this powerful strategy that exploits advanced bound-state solvers to access continuum physics. The choice then becomes the method for inferring the real-$E$ results from the complex-$E$ solutions.

To do so, the CE method uses continued-fraction based  extrapolants~\cite{Schlessinger:1966zz, Schlessinger:1968vsk, Schlessinger:1968zz, McDonald:1969zza, Uzu:2003ms, Deltuva:2012fa, Deltuva:2013qf,Deltuva:2013mda,Deltuva:2014pda} while LIT fits real-$E$ observables via integral transforms~\cite{efros1985computation, Efros:1994iq, Efros:2007nq,Orlandini:2013eya,Sobczyk:2021dwm,Sobczyk:2023sxh,Bonaiti:2024fft}. In contrast, our method uses the RBM emulator to realize this connection, which keeps the number of training calculations minimal and offers extreme interpolating and extrapolating speed. Interestingly, this work creates an NHQM approach that computes both 
 observables and spectra and obtains an insight that the complex-energy based methods belong to the type-II category mentioned earlier. 

Importantly, our $\vectheta$-emulation, indispensable for exploring parameter space, can be adapted to work with the CE and LIT methods. It provides fast \emph{interpolations} of $\resolventamp$ at general $\vectheta$ and complex energies (close to $E_\alpha^\mathrm{tr}$), which can be used as the inputs for the complex-$E$-to-real-$E$ procedures in CE and LIT, creating fast $\vectheta$-emulations of their continuum calculations. See Ref.~\cite{Zhang:2024gac} for details. 

Moreover, this work expands the MOR literature by applying the RBM method to project an operator with a continuous spectrum into a finite matrix; previous MOR literature~\cite{AntoulasBook2005} mainly focuses on dimension reduction of large matrices. Our study also shows great similarities between the complex-$E$ emulation and the ORA for functions with branch points~\cite{Nakatsukasa_2018, Trefethen2021NodeClustering, Trefethen2023review, nakatsukasa2023years}. 

\paragraph{Further studies.} 
One direction is to implement greedy algorithms~\cite{Sarkar:2021fpz,Maldonado:2025ftg} \cor{for selecting training points (including with random $\Im E^\mathrm{tr}_\alpha$)}, as suggested by the redundancy ($\Neff \ll N_b$) observed above. \cor{We then obtain optimal bases with minimal training costs.} \cor{The algorithms} would also provide a more sophisticated regularization than the current SVD truncation, further stabilizing the analytical continuation in the complex-$E$ plane as seen in the context of ORA~\cite{Nakatsukasa_2018}. The \cor{continuation, in particular near thresholds, could be also enhanced by enforcing general analytical behaviors in $\matrixform{\resolventamp}$ within, e.g., the framework of Bayesian statistics~\cite{Hicks:2022ovs}}. Lastly, we can extend the framework to treat systems with strong Coulomb effects by employing Coulomb wave functions~\cite{Efros:2007nq}.

\paragraph{Summary.}
We apply the RBM method to emulate inhomogeneous \schro equations in a combined space of the complex-$E$ plane and other parameters in the equations. It creates an NHQM method for extracting and emulating continuum states based on bound-state-like calculations. However, the method differs from previous NHQM methods in that it constructs the subspace by RBM-based projections, while the previous methods form a many-body basis with direct products of single-particle states. Good emulation performances are demonstrated in two and three-body systems as proof of principles. 
The method can also help emulate existing continuum calculations, such as those based on CE and LIT methods. Finally, the potential connection between our emulation and ORA should be further explored.

\paragraph{Acknowledgements.}
Discussions with Dean Lee, Dick Furnstahl, Nobuo Hinohara, Chong Qi, Simin Wang, Alex Gnech, Bijaya Acharya, and Ante Ravlic are greatly appreciated. I also thank  Dick Furnstahl for careful reading of the manuscript. This material is based upon work supported by the U.S. Department of Energy, Office of Science, Office of Nuclear Physics, under the FRIB Theory Alliance Award No. DE-SC0013617 and under the  STREAMLINE Collaboration Award No. DE-SC0024586 (Michigan State University).



%

\end{document}